\documentclass[pra,eqsecnum,showpacs,twocolumn]{revtex4}
\usepackage{newlfont}
\usepackage{amssymb}
\usepackage{amsfonts}
\usepackage{amsmath}
\usepackage{graphicx}
\usepackage{bm}
\begin{document}
\title{State estimation on correlated copies}
\author{Rafa{\l} Demkowicz-Dobrza\'nski}
\affiliation{Center for Theoretical Physics, Polish Academy of Sciences\\
Aleja Lotnik{\'o}w 32/44, 02-668 Warszawa, Poland}
\date{\today}
\begin{abstract}
State estimation is usually analyzed in the situation when copies are in a product state, either mixed or pure. 
We investigate here the concept of state estimation on correlated copies. 
We analyze state estimation 
on correlated $N$ qubit states, which are permutationally invariant.
Using a correlated state we try to estimate as good as possible the direction of the Bloch vector of a single particle 
reduced density matrix. We derive the optimal fidelity for all permutation invariant states. We find the optimal state, 
which yields the highest estimation fidelity among the states with the same reduced density matrix. 
Interestingly this state is not a product state. We also point out that states produced by optimal universal cloning machines
are the worst from the point of view of estimating the reduced density matrix.
\end{abstract}
\pacs{03.65.Ta, 03.67.Hk}

\maketitle

\section{Introduction}
In the standard approach to quantum state estimation, one is given a state which is a product state of $N$ identical copies of a quantum system in an unknown quantum state.
The task is to find the optimal measurements on the $N$ copies and the optimal guessing strategies in order to 
learn what is the actual state of the quantum system. Optimal strategies differ depending on many factors: What figure of merit describes the accuracy of the guess?
What is the dimensionality of the Hilbert space of a single copy? How many copies of a state are given? What is the prior knowledge about the state (in particular is it mixed or pure)? Is there a limitation on the allowed measurements (for example restriction to local measurements - separate measurement on each copy)?

Many results have been obtained in recent years on the subject. In particular the optimal fidelity for estimating the state of $N$ qubits in an unknown pure state
was found in \cite{massar1993}. More generally, optimal fidelity for estimating arbitrary pure state of a $d$-level system was derived in \cite{bruss1999,hayashi2004}.
The case of mixed states is more demanding, and the general results for the optimal estimation fidelity are known only for the case of qubits
\cite{vidal1999,cirac1999}. Estimation via local measurements was analyzed in the case of qubits for pure and mixed states \cite{bagan2002,bagan2004}.

In all the references mentioned above, state estimation is performed on the input state of the form: $\rho^{\otimes N}$, where $\rho$ is either pure or mixed state of a single copy.
This is a product state in which copies are not correlated. The intriguing questions is: what is the optimal state estimation when copies are correlated? 
Do correlated copies allow for a better estimation of the reduced density matrix, or on the contrary, do correlations diminish the estimation fidelity? 
In other words is the product state $\rho^{\otimes N}$ optimal from the point of view of state estimation?

In this paper we analyze the state estimation on correlated qubits. 
We analyze correlated density matrices of $N$ qubits, which are permutationally invariant, and have single particle
density matrices $\rho$, described by a Bloch vector of length ${r}$. 
We pose the problem of the optimal estimation of the direction of the Bloch vector given the correlated state of $N$ qubits.
We find out that the product state $\rho^{\otimes N}$ is not the optimal state for the purpose of state estimation.
We derive the optimal state, and the corresponding \emph{optimal fidelity of estimation of the optimal state}. 
Additionally we also point out the fact that the state coming out from the optimal cloning machine is in fact the worst state from the point of view of state estimation, 
among all the states with the same single particle reduced density matrix.

\section{Cloning and state estimation - the motivation}
There is an interesting connection between optimal cloning and optimal state estimation, which appeared to be fruitful in deriving optimal fidelity for $N \to M$ 
cloning of qubits, using the known result for state estimation \cite{bruss1997}. Interestingly, benefits flew also in the opposite direction, as knowledge of the optimal
local cloning fidelity of a pure state of a $d$-level system\cite{werner1998} allowed the derivation of the optimal estimation fidelity in this case \cite{bruss1999}.
The main result relating cloning and estimation is, that the local fidelity of $N \to \infty$ optimal cloning of a quantum state is equal to the fidelity of 
optimal estimating the state using $N$ copies.

In the derivation of the above result, as one of the steps in the proof,  the state estimation on the clones coming out of the cloning machine is considered. 
Clones coming out from the optimal $N \to M$ cloning machine are always correlated (with the exception of the trivial case $N=M$). This is an example of the situation 
where one has to consider state estimation on correlated copies. For the sake of the proof authors make use of the fact that the state of the clones is supported on
the symmetric subspace, and thus can be decomposed as \cite{werner1998}
\begin{equation}
  \label{eq:11}
  \tilde{\rho} = \sum_i \alpha_i |\psi_i \rangle \langle \psi_i|^{\otimes N},
\end{equation}
where $\sum_i \alpha_i =1$ and $\alpha_i$ can be both positive or negative (from now on $\tilde{\rho}$ denotes 
the full density matrix, while $\rho$ denotes single particle reduced density matrix).
This decomposition allows one to relate the fidelity of state estimation on correlated clones 
to the fidelity of estimation on uncorrelated pure states \cite{bruss1997,bruss1999}. 

Some additional observations can be made concerning the relation between cloning and state estimation. Consider optimal universal $N \to M$ cloning of qubits.
If the input $N$ copies were in the state $|\psi\rangle^{\otimes N}$ then the single-particle reduced density matrix of each clone is given by \cite{gisin1997,bruss1997}:
\begin{equation}
  \label{eq:1}
  \rho_{\mbox{\tiny{$N\to M$}}} = {r}(N,M)|\psi\rangle \langle \psi| + \frac{1-{r}(N,M)}{2}\openone,
\end{equation}
where 
\begin{equation}
  \label{eq:2}
  {r}(N,M) = \frac{N}{M}\frac{M+2}{N+2}.
\end{equation}
When the number of clones $M$ goes to $\infty$ the reduced density matrix of each clone reads:
\begin{equation}
  \label{eq:3}
  \rho_{\mbox{\tiny{$N\to\infty$}}}=\frac{N}{N+2}|\psi\rangle \langle \psi|+\frac{1}{N+2}\openone.
\end{equation}
Estimation of the initial state $|\psi\rangle$ can be done using these infinite number of clones (in reality a very large number), and the resulting 
fidelity is \cite{bruss1997}:
\begin{equation}
  \label{eq:4}
  F=\frac{N+1}{N+2}.
\end{equation}
This result is obtained using the fact that the full density matrix  of clones is supported on symmetric subspace, and this fidelity is just the fidelity 
of optimal state estimation on $N$ qubits. 

Let us assume now (counter-factually) that after $N \to \infty$ cloning, the reduced density matrix of each clone is given by the same formula as in Eq.~(\ref{eq:3}),
but unlike in real cloning machine, the full density matrix of clones is a product state: $\tilde{\rho}_\textrm{prod}=\left(\rho_{\mbox{\tiny{$N\to\infty$}}}\right)^{\otimes \infty}$.
What is the fidelity of estimating the initial state $|\psi\rangle$ using the state $\tilde{\rho}_\textrm{prod}$. Each single-particle density matrix, has a Bloch vector 
pointing in the direction $|\psi\rangle$.
Since we have infinite number of such density matrices independent one from another, with arbitrary high accuracy we can estimate the direction of the Bloch vector, 
and consequently
the state $|\psi\rangle$. The fidelity of estimating $|\psi\rangle$ using the state $\tilde{\rho}_\textrm{prod}$ is thus $1$.
Consequently, if clones were really in such a product state we could beat the state estimation fidelity 
limit.
This observation shows, that the restrictions on cloning coming from the state estimation are not confined to the restriction on local fidelity of cloning. 
State estimation imposes also some necessary degree of correlation between the clones. 

In this context, it is interesting, to study in general the state estimation fidelity on correlated copies. One could then impose limitations on possible states of clones solely
using limits from state estimation theory.

The other problem, on which the analysis of estimation on correlated copies could shed some light, is retrieving information from
quantum systems after they have passed through a noisy channel. Consider for example $N$ qubits in a pure state $|\psi \rangle$ 
that pass through a noisy channel, and each qubit at the output has reduced density matrix $\rho$, with the Bloch vector 
shrunk by certain factor. One can ask the question, what kind of noise leads to smallest (largest) loss of information about
initial state $|\psi \rangle$? Is the noise acting independently on qubits less harmful than the one inducing correlation between them?

\section{State estimation on correlated copies}
The problem of state estimation on correlated copies can be stated as follows. Consider a density matrix  $\tilde{\rho}$ describing the state of $N$
$d$-level systems. In order not to distinguish any of the systems we shall assume that $\tilde{\rho}$ is permutation invariant i.e.:
\begin{equation}
  \label{eq:18}
  \Pi\tilde{\rho}\Pi^\dagger = \tilde{\rho},
\end{equation}
where $\Pi$ is an arbitrary permutation of $N$ systems. 
Single particle reduced density matrices of each system are equal and denoted by $\rho=\textrm{Tr}_{2\dots N}\tilde{\rho}$. 
The goal is to find optimal measurements on the state $\tilde{\rho}$ and optimal guessing strategies in order to estimate the state $\rho$.
The measurement is described by a positive operator values measure \cite{helstrom1976}: $\sum_\mu P_\mu = \openone$. With each measurement result a guess $\rho_\mu$ is associated.
If $F\left(\rho_\mu,\rho\right)$ denotes a figure of merit, describing how close a guess $\rho_\mu$ is to the actual single particle state $\rho$, then 
the average fidelity of the state estimation is given by the formula:
\begin{equation}
  \label{eq:5}
  F = \overline{\sum_\mu F\left(\rho_\mu,\rho\right) \textrm{Tr}\left(P_\mu \tilde{\rho}  \right)},
\end{equation}
where averaging is performed over the ensemble of unknown initial states $\tilde{\rho}$. 

\section{Optimal state estimation on a correlated N qubit state}
\label{optimal:estimation}
We now move on to the specific case of state estimation on a correlated state of $N$ qubits. 

Single particle density matrix of each copy
is given by:
\begin{equation}
  \label{eq:6}
  \rho = {r}|\psi\rangle \langle \psi| + \frac{1-{r}}{2}\openone = \frac{1}{2}\left(\openone + \vec{\sigma} \vec{r} \right).
\end{equation}
We assume that the length ${r}$ of the Bloch vector $\vec{r}$ is fixed and our goal is to estimate its direction (this is for example the case 
when relating cloning with state estimation).
In other words we want to estimate the pure state $|\psi\rangle$, given a mixed state $\tilde{\rho}$ with single particle reduced density matrices $\rho$.
In what follows we shall assume that we have no prior knowledge about the direction of the single particle Bloch vector.
Averaging in Eq.~(\ref{eq:5}) will thus be performed over the ensemble of input states of the form: 
\begin{equation}
\label{eq:ensemble}
\tilde{\rho}(U)=U^{\otimes N} \tilde{\rho} U^{\dagger \otimes N}
\end{equation}, 
where $U$ are elements of the $\textrm{SU}(2)$ defining representation, distributed uniformly with respect to the invariant measure on the $\textrm{SU}(2)$ group. 

As a figure of merit we shall adopt the fidelity: $F(\rho_\mu, |\psi\rangle \langle\psi|)= \langle \psi| \rho_\mu |\psi \rangle$.
If the full density matrix is a product one: $\tilde{\rho}_\textrm{prod}=\rho^{\otimes N}$ then the optimal fidelity of state estimation given in \cite{cirac1999} can be rewritten as:
\begin{widetext}
\begin{equation}
  \label{eq:7}
F_{\textrm{prod}}=\sum\limits_{j=s}^{N/2} \frac{d_j \left(1-{r}^2\right)^{N/2-j}}{2^{N+3}{r}^2(j+1)} \left[(1+{r})^{2j+1}\left[{r}(3+4j)-1 \right]+(1-{r})^{2+2j}\right],
\end{equation}
\end{widetext}
where $d_j$ (for $j \neq N/2$) is defined as:
\begin{equation}
  \label{eq:1a}
  d_j = \left( \begin{array}{c} N \\ \frac{N}{2}-j \end{array} \right) - \left( \begin{array}{c} N \\ \frac{N}{2}-j-1 \end{array} \right),
\end{equation}
$d_{N/2}=1$, and  $s=0\ (1/2)$ for $N$ even (odd).

Instead of the product state, consider now the state $\tilde{\rho}_\textrm{sym}$, which is supported on the symmetric subspace (for example this is the case of the clones coming out 
from the optimal cloning machine), and has the same single particle reduced density matrix $\rho$.
The two states $\tilde{\rho}_\textrm{prod}$, $\tilde{\rho}_\textrm{sym}$ coincide only when ${r}=1$ (pure state case).
The optimal fidelity of estimating the direction of the Bloch vector $\vec{r}$ (\ref{eq:6}) reads:
\begin{equation}
  \label{eq:8}
  F_\textrm{sym} = \frac{1}{2}\left(1+{r}\frac{N}{N+2}\right).
\end{equation}
This result reveals what is known from analyzing cloning in stages, namely that shrinking factors multiply \cite{bruss1997}. Note that here we abstract from 
the problem of cloning, and analyze general state $\rho_\textrm{sym}$ supported on symmetric subspace, which may be 
not be obtainable by means of optimal cloning machines.
The proof of the above formula, using the methods from \cite{bruss1997,cirac1999},   
goes as follows. 
The fidelity of estimation can be written as:
\begin{equation}
  \label{eq:13}
  \begin{array}{rcl}
  F&=& \overline{\langle \psi| \sum_\mu \rho_\mu \textrm{Tr}\left(P_\mu \tilde{\rho}\right)||\psi \rangle} = \\
  \\
   &=& \overline{\langle\psi| \rho_\textrm{est} |\psi \rangle},
  \end{array}
\end{equation}
where $\rho_\textrm{est}$ is the state we estimate on average (average with respect to different measurement results $\mu$). 
The state estimation procedure may thus be described by a completely positive map $\mathcal{E}$:
\begin{equation}
  \label{eq:14}
 \rho_\textrm{est}=\mathcal{E}(\tilde{\rho}), \quad F=\overline{\langle \psi| \mathcal{E}(\tilde{\rho})|\psi\rangle}.
\end{equation}
Thanks to the choice of the ensemble of states over which we average, the search for the optimal state estimation procedures 
can be restricted to the set of covariant operations $\mathcal{E}$
\cite{cirac1999} i.e.:
\begin{equation}
  \label{eq:15}
  \mathcal{E}\left(U^{\otimes N} \tilde{\rho} U^{\dagger \otimes N}\right) = U \mathcal{E}\left(\tilde{\rho}\right) U^\dagger.
\end{equation}
Covariance property of $\mathcal{E}$ implies that the action of $\mathcal{E}$ on a pure product state $|\phi\rangle \langle \phi|^{\otimes N}$ yields:
\begin{equation}
  \label{eq:16}
  \mathcal{E}\left(|\phi\rangle \langle \phi|^{\otimes N}\right)= a|\phi \rangle \langle \phi|+\frac{1-a}{2}\openone,
\end{equation}
where $a \geq 0$ and does not depend on $|\phi\rangle$ (this is because the result must commute with any rotation around the 
axis defined by the Bloch vector of $\phi$).

Using the decomposition (\ref{eq:11}) of the matrix $\tilde{\rho}$, and the fact that
\begin{equation}
  \label{eq:20}
  \rho=\textrm{Tr}_{2\dots N}(\tilde{\rho})=\sum_i \alpha_i |\psi_i \rangle \langle \psi_i|,
\end{equation}
the fidelity of state estimation can be written as (averaging can be dropped out thanks to covariance):
\begin{equation}
  \label{eq:12}
  \begin{array}{rcl}
    F &=& \langle \psi| \mathcal{E}\left(\sum\limits_i \alpha_i |\psi_i\rangle\langle \psi_i|^{\otimes N}\right) |\psi \rangle  = \\
&=& \langle \psi|\sum\limits_i \alpha_i \mathcal{E}\left(|\psi_i\rangle\langle \psi_i|^{\otimes N}  \right) |\psi \rangle = \\
&=& \langle \psi|\sum\limits_i \alpha_i \left(a|\psi_i \rangle \langle \psi_i|+\dfrac{1-a}{2}\openone\right)|\psi\rangle= \\
&=& \langle \psi|\left[ a \left( {r} |\psi\rangle \langle \psi| + \dfrac{1-{r}}{2}\openone \right) +\dfrac{1-a}{2}\openone\right] |\psi\rangle= \dfrac{a {r} +1}{2}.
  \end{array}
\end{equation}
The highest fidelity corresponds to the highest possible value of $a$. From Eq.~(\ref{eq:16}) one sees that the highest possible $a$  corresponds to the estimation strategy optimal
for pure product states i.e. $a=N/(N+2)$\cite{massar1993}. Substituting this value into the Eq.~(\ref{eq:12}) one gets (\ref{eq:8}). $\blacksquare$

Interestingly the relation  $F_\textrm{prod} \geq F_\textrm{sym}$ holds, with equality only when ${r}=0$ or ${r}=1$.
This may not seem obvious from the formulas (\ref{eq:7}),(\ref{eq:8}), but will become clear in Sec.~\ref{sec:optimalstate}. 

We now derive the optimal estimation fidelity for a more general class of density matrices, namely when $\tilde{\rho}$ is an arbitrary permutation invariant density matrix.
As a byproduct of the derivation we will be able to write down the state
$\tilde{\rho}_\textrm{opt}$, which allows for the best estimation of the direction of the single particle Bloch vector, among all the density matrices 
$\tilde{\rho}$ with fixed single particle reduced density matrix $\rho$. Interestingly this in general will not be the product state $\tilde{\rho}_\textrm{prod}$.

We shall follow the methods presented in \cite{cirac1999}. In the space of $N$ qubits (spin $1/2$) one can introduce an orthonormal basis $|j,m,\alpha\rangle$ 
[$j=s, \dots, N/2$ denotes the total spin, whin  $s= 0\ (1/2)$ for $N$ even (odd)] 
such that: (i) for a fixed $j$ and  $\alpha$ vectors $|j,m,\alpha\rangle$ span the ($2j+1$) dimensional space, in which an irreducible representation of $\textrm{SU}(2)$ acts.
(ii) for a fixed $j$ and $m$ vectors $|j,m,\alpha\rangle$ form a basis for $d_j$ dimensional (\ref{eq:1a}) irreducible representation of the symmetric group ($S_N$).
This representation correspond to a Young diagram with two rows of lengths $N-2j$ and $2j$. 
Let $U_{j,\alpha}$ be unitary matrices that connect spaces with fixed $(j, \alpha)$ with the space $(j, 1)$: $|j, m, 1\rangle =
U_{j,\alpha}|j, m, \alpha\rangle$.
For convenience it is chosen that: $|j,m,1\rangle = |j,m \rangle \otimes |\Psi_-\rangle^{\otimes N/2-j}$, where $|j,m\rangle$ is the symmetric state 
of $2j$ qubits with the total spin $j$ and the projection of the spin onto the $\vec{r}$ direction (direction of the single particle Bloch vector) 
equal $m$, while $|\Psi_-\rangle$ is the singlet state.

The above decomposition of the Hibert space into subspaces invariant under permutations and $\textrm{SU}(2)^{\otimes N}$ 
is a manifestation of the Schur-Weyl theorem \cite{fulton1991}. 
The symmetric group acts irreducibly in the subspaces with fixed $j$ and $m$.
This allows us to write down the most general permutationally invariant density matrix $\tilde{\rho}$, in the similar way that
the product state $\rho^{\otimes N}$ was written down in \cite{cirac1999}:
\begin{equation}
  \label{eq:9}
  \tilde{\rho} = \sum\limits_{j=s}^{N/2} p_j \frac{1}{d_j} \sum\limits_{\alpha=1}^{d_j} \tilde{\rho}_{j,\alpha},
\end{equation}
where
\begin{equation}
  \label{eq:10}
  \begin{array}{rcl}
  \tilde{\rho}_{j,\alpha} &=& U_{j,\alpha}\tilde{\rho}_{j,1}U^\dagger_{j,\alpha}\\
  \\
  \tilde{\rho}_{j,1}&=& \tilde{\rho}_j \otimes |\Psi_-\rangle \langle \Psi_-|^{\otimes N/2-j} \\
  \\
  \tilde{\rho}_j&=&\sum\limits_{m, m^\prime=-j}^j \lambda_{j, m, m^\prime}|j,m^\prime\rangle \langle j,m|.
\end{array}
\end{equation}
Coefficients $p_j$ are nonnegative and satisfy  $\sum_{j=s}^{N/2} p_j = 1$, while for a fixed $j$ the matrix $\lambda_{j,m,m^\prime}$ is a positive semi-definite matrix with trace one.
Summation over $\alpha$ assures that the density matrix $\tilde{\rho}$ will indeed be permutation invariant.

Let us now express the single particle reduced density matrix $\rho$ with the help of coefficients $p_j, \lambda_{j, m, m^\prime}$. 
First we calculate the single particle reduced density matrix of the symmetric matrix $\tilde{\rho}_j$. Written in the basis $|\psi\rangle$, $|\psi\rangle^\perp$ it reads:

\begin{equation}
  \label{eq:17}
   \rho_j = \textrm{Tr}_{2,\dots,2j}\left(\tilde{\rho}_j\right)= 
   \left(\begin{array}{cc} 
       A_j & C_j  \\ 
C_j^*  & B_j,
   \end{array}\right)
\end{equation}
where
\begin{align}
A_j=\sum_{m=-j}^j \frac{j+m}{2j} \lambda_{j, m, m},\quad B_j=\sum_{m=-j}^j \frac{j-m}{2j}\lambda_{j, m, m}  \notag\\
C_j=\sum_{m=-j+1}^j \frac{1}{2j}\sqrt{(j+m)(j-m+1)}\lambda_{j, m, m-1} \notag
\end{align}
The Bloch vector corresponding to the density matrix $\rho_j$ reads:
\begin{equation}
\vec{r}_j=\left[C_j+C_j^*,i(C_j-C_j^*),A_j-B_j\right]
\end{equation}
and its length is equal to:
\begin{equation}
{r}_j=|\vec{r}_j|=\sqrt{(A_j-B_j)^2-2|C_j|^2}
\end{equation}

Using formulas (\ref{eq:9}),(\ref{eq:10}), we obtain the single particle reduced density matrix $\rho$:
\begin{equation}
  \label{eq:19}
  \rho=\textrm{Tr}_{2,\dots,N}(\tilde\rho)= \sum\limits_{j=s}^{N/2} p_j \left(\frac{2j}{N}\rho_j+\frac{N-2j}{2N}\openone \right) 
\end{equation}
Comparing it with the single particle reduced density matrix (\ref{eq:6}), one arrives at the following relation between
local Bloch vector $\vec{r}$ and vectors $\vec{r}_j$:
\begin{equation}
\label{eq:nvec}
\vec{r}=\frac{2}{N}\sum_{j=s}^{N/2} p_j j \vec{r}_j.
\end{equation}
This condition guarantees that the state $\tilde{\rho}$, parameterized by $p_i$, and $\lambda_{i,m,m^\prime}$, yields
the single qubit reduced density matrix with the Bloch vector equal to $\vec{r}$.

\subsection{Optimal estimation strategy}

We are now prepared to derive the fidelity of the optimal state estimation performed on the state $\tilde{\rho}$ parameterized by $p_i, \lambda_{j,m, m^\prime}$.
First let us observe that similarly as in \cite{cirac1999,bagan2006a}, we do not loose optimality if we start with projecting the state $\tilde{\rho}$ on
a subspace with given $(j, \alpha)$. This is simply a measurement projecting on subspaces with respect to which $\tilde{\rho}$ is already block diagonal, and therefore
this operations cannot decrease accessible information on the state.
If the measurement yields  $\alpha\neq1$ then $U_{j,\alpha}$ is applied in order to move the state to the subspace $(j,1)$, and consequently move singlets to the last $N/2-j$ qubits.
Since singlets yield no information on the direction of the local Bloch vector they can be discarded.

After this preprocessing, with probability $p_j$, we have at our disposal a $2j$ qubit state $\tilde{\rho}_j$. Further procedure depends on
our knowledge on the structure of correlations in the state $\tilde{\rho}$. 

\textbf{Known structure of correlations.}
Knowing the structure of correlations means that 
we know everything about the state $\tilde{\rho}$ except for its possible local rotation with $U^{\otimes N}$. This means that even though we do not 
know the direction of the local Bloch vector $\vec{r}$, we know the relative orientation of vectors $\vec{r}_j$ in 
Eq.~(\ref{eq:nvec}). In other words, we know all scalar products between $\vec{r}$ and $\vec{r}_j.$

This information can be helpful in increasing our fidelity of estimation. 
The state $\tilde{\rho}_j$ which we obtain from the preprocessing stage is a $2j$ qubit state supported on the fully symmetric space.
We will use the estimation of the direction of its local Bloch vector $\vec{r}_j$ 
(optimal estimation of symmetric state described at the beginning of this section)
 as an information about the direction $\vec{r}$.  We may ask if it is advantageous to make 
additional transformation of our guesses about the direction $\vec{r}_j$ in order to make them
closer to the vector $\vec{r}$ (higher overlap) and hence increase the fidelity. Note that we do not know the direction
$\vec{r}$ so this transformation has to be defined independently of $\vec{r}$.
 
We know the scalar product $\vec{r}_j\cdot \vec{r}$, which tells us that if the direction $\vec{r}_j$ is set to point to the
north pole then $\vec{r}$ lies somewhere at a parallel of colatitude 
$\theta_j=\arccos(\vec{r}_j \cdot \vec{r})$. Since we completely do not know
at which point on the parallel lies the vector $\vec{r}$, then thanks to the rotational symmetry 
of the problem, it may shown that if $\theta_j \leq 90^\circ$ any transformation of guesses of $\vec{r}_j$ will actually lead to
a lower average overlap -- it is optimal to keep the guesses as if we wanted to guess vector $\vec{r}_j$ at the north pole. 
If, however, $\theta_j > 90^\circ$ then it is clearly preferable to make a reflection with respect to the center of the sphere
as if we wanted to estimate the vector pointing to the south pole. 

Consequently, the optimal way to use the state $\tilde{\rho}_j$ for determining the direction $\vec{r}$, 
is to estimate the direction of its local Bloch vector $\vec{r}_j$ and in the case
$\theta_j > 90^\circ$  make an additional reflection of the guesses 
\footnote{See also \cite{bagan2006a}, for a strict derivation of the optimal estimation strategy, where although the estimation is considered only on product states $\rho^{\otimes N}$, the methods can be applied to the correlated case also, since they rely mainly on permutational invariance of the state}.

We can introduce $\rho_j^\prime$ which is equal to the reduced density matrix $\rho_j$ of the state $\tilde{\rho}_j$, 
whenever $\theta \leq 90^\circ$, and when $\theta > 90^\circ$ its Bloch vector is additionally reflected.
The optimal fidelity of estimating the direction of $\vec{r}$ 
using the state $\tilde{\rho}_j$ reads (compare the derivation (\ref{eq:12})):
\begin{align}
F_j &= \langle \psi| \left(\frac{2j}{2j+2}\rho_j^\prime +\frac{1}{2j+2}\openone \right)| \psi \rangle= \notag \\ 
&=\frac{j}{j+1}\max(A_j,B_j)+\frac{1}{2j+2},
\end{align} 
where $|\psi \rangle$ is a pure state with the normalized Bloch vector $\vec{r}/{r}$ and $\max(A_j,B_j)$ appears 
as a result of the conditional reflection of the Bloch vector of $\rho_j^\prime$.

Expressing $F_j$ explicitely using Bloch vectors, and summing over $j$ with weights $p_j$ we get the fidelity of estimation when
the structure of correlations is known:
\begin{equation}
\label{eq:fidknown}
F_\textrm{known}=\frac{1}{2}\left(1 + \sum_{j=s}^{N/2} \frac{p_j j}{j+1} \frac{|\vec{r}_j \cdot \vec{r}|}{{r}} \right).
\end{equation}  

\textbf{Unknown structure of correlations.}
In this case we do not know anything about the structure of correlations in the state $\tilde{\rho}$. In particular we do not know relative
orientations of vectors $\vec{r}_j$, and cannot judge whether $\theta_j\leq 90^\circ$ or $\theta_j > 90^\circ$.
As a result we do not know when to apply reflection, and the natural \footnote{We use the word ``natural'' instead of ``optimal'' here,
not because we do not know the optimal strategy, but because in this setting the notion of optimality is vague.
Since we assume here some additional ignorance, apart from the action of unknown local unitaries $U^{\otimes}$, then in order to speak about
optimality we should describe precisely the prior distribution of different structures of correlations. This would make
the notion of ``unknown'' structure of correlation strict. We do not see the natural choice for this distribution,
apart from the observation that in any ``natural'' choice, the equation Eq.~(\ref{eq:nvec})
suggests that average vectors $\vec{r}_j$ will more likely point in the same direction as $\vec{r}$ than in the opposite one.}
approach in this case is not to apply any reflection at all. 
Therefore, we estimate the direction of $\vec{r}_j$ using the state $\tilde{\rho}_j$,
and regard this result as an estimate of the direction of $\vec{r}$.

The only difference in the formula for the fidelity as compared to Eq.~(\ref{eq:fidknown}) is the lack of the absolute value,
since we do not apply reflection if $\theta_j > 90^\circ$. The fidelity reads: 
\begin{equation}
\label{eq:fidunknown}
F_\textrm{unknown}=\frac{1}{2}\left(1 + \sum_{j=s}^{N/2} \frac{p_j j}{j+1} \frac{\vec{r}_j \cdot \vec{r}}{{r}} \right)
\end{equation}  
Obviously, in general we will arrive at a lower estimation fidelity than the one given in Eq.~(\ref{eq:fidknown}). 

\section{$N$ qubit state optimal for state estimation} 
\label{sec:optimalstate}
We are now able to answer the question posed at the beginning of this paper. With a fixed length of the Bloch vector ${r}$
of a single particle reduced density matrix, what is the state $\tilde{\rho}$ optimal for estimation of the direction of the Bloch vector?
Mathematically the goal is to optimize estimation fidelity $F_\textrm{known}$ or $F_\textrm{unknown}$ under the constraints:
\begin{align}
\label{eq:constraint1}
&\frac{2}{N}\sum_{j=s}^{N/2} p_j j \vec{r}_j = \vec{r}\\
\label{eq:constraint2}
&\sum_{j=s}^{N/2} p_j = 1, \quad 
p_j \geq 0, \quad {r}_j \leq 1
\end{align}

Since $F_\textrm{known}$ or $F_\textrm{unknown}$ depend on $\vec{r}_j$ only via scalar products $\vec{r}_j \cdot \vec{r}$,
we can always replace vectors $\vec{r}_j$, by their projections
 on the direction $\vec{r}$. Both the constraint (\ref{eq:constraint1}) and the fidelities are not affected by this. 
From now on, we can thus regard all vectors $\vec{r}_j$ lying in the same line as $\vec{r}$, and the constraint
(\ref{eq:constraint1}) reads:
\begin{equation}
\label{eq:constraint3}
\frac{2}{N}\sum_{j=s}^{N/2}\pm  p_j j r_j = r\\
\end{equation}
where $\pm$ corresponds to vector $\vec{r}_j$ being parallel and antiparallel to $\vec{r}$ respectively.

Let us start with the case when structure of correlations is known.

\textbf{Known structure of correlations.}
Looking at Eq.~(\ref{eq:fidknown}), we notice that it is always advantageous for the value of fidelity to take maximal possible length of vectors $\vec{r}_j$, i.e. 
${r}_j=1$. We can always do it, and still keep the constraint (\ref{eq:constraint3}) satisfied, since if necessary we can choose some vectors 
$\vec{r}_j$ to be antiparallel, and change weights $p_j$ appropriately. Consequently, we have to maximize the quantity (note $|\vec{r}_j \cdot \vec{r}|/r=1$):
\begin{equation}
\label{eq:delta}
\Delta_\textrm{known}=\sum\limits_{j=s}^{N/2} \dfrac{p_j j}{j+1}
\end{equation}
Notice that $j/(j+1)$ is an increasing function for positive $j$. We should therfore try to give as much weight as possible to $p_j$ with the highest $j$. 
Ideally we would like to have $p_{N/2}=1$ and all other $p_j=0$. This, however, is only possible when ${r}=1$. In other cases the condition (\ref{eq:constraint1}) would be
violated. If ${r}<1$ we need to include some other $p_j$, the optimal choice is to take additionally only the second highest, i.e. with $j=N/2-1$. 
In order to satisfy the condition (\ref{eq:constraint1}), and maximize $\Delta$ the direction of $\vec{r}_{N/2}$, $\vec{r}_{N/2-1}$ should 
be respectively parallel and antiparallel to $\vec{r}$ (in terms of $\lambda_{j,m,m^\prime}$ coefficients, this means
that $\lambda_{N/2,N/2,N/2}=1$, $\lambda_{N/2-1,-(N/2-1),-(N/2-1)}=1$, and all 
other coefficients are zero). 

Constraint (\ref{eq:constraint1}) implies that
\begin{equation}
\label{eq:subtr}
\frac{2}{N}\left(p_{N/2}N/2-p_{N/2-1}\left(N/2-1\right)\right)={r}. 
\end{equation}
Together with the fact that $p_{N/2}+p_{N/2-1}=1$ we get
explicitely:
\begin{displaymath}
\left\{
\begin{array}{lcl}
p_{N/2}&=& \dfrac{N({r}+1)-2}{2N-2}   \\
p_{N/2-1}&=& \dfrac{N-{r}N}{2N-2} \quad \textrm{(antiparallel)}
\end{array}
\right.
\end{displaymath}

and the corresponding estimation fidelity reads:
\begin{equation}
F=\frac{N^2+{r}-2}{(N-1)(N+2)}.
\end{equation}
The fidelity depends very weakly on ${r}$, which means that even if we have very mixed local density matrices -- very short local Bloch vector $\vec{r}$ -- we can have 
a very good estimation of the direction $\vec{r}$. This can be easily understood, when realizing that the short Bloch vector $\vec{r}$ arises from subtraction
of two long vectors (see Eq.~(\ref{eq:subtr})). Moreover, we can perfectly distinguish subspaces with $j=N/2$ and $j=N/2-1$.
Hence, when we measure $j=N/2-1$ we know that $\vec{r}_{N/2-1}$ points in the opposite direction than $\vec{r}$, and we reflect it. As a result
we get very good estimation of the vector $\vec{r}$. In the asymptotic limit $N \to \infty$, up to the leading order in $1/N$ the fidelity reads: 
$F=1-1/N$. It does not depend on ${r}$ and is the same as the asymptotic fidelity when estimating $N$ copies of a pure state. 

For this strategy to be applicable, we need to perform collective measurements. If we restrict ourselves to local measurements assisted with 
classical communication, we can not discriminate between spaces with different $j$ with high confidence.
This leads to an interesting situation 
when global measurements allow for almost perfect estimation of direction, while local measurement, even in a limit of large number of copies yield very poor
estimation quality \cite{bagan2006}.

\textbf{Unknown structure of correlations.}
Our goal is to find a state that under the fixed length ${r}$ of the local Bloch vector will maximize the fidelity in Eq.~(\ref{eq:fidunknown}).
As a matter of fact we need to maximize:
 \begin{align}
 \label{eq:delta2}
 \Delta_\textrm{unknown}=\left(\sum_{j=s}^{N/2} \frac{\pm p_j j {r}_j}{j+1}\right) 
 \end{align}
where $\pm$ indicates whether the vector $\vec{r}_j$ is parallel or antiparallel to $\vec{r}$. The presence of the $\pm$ sign in Eq.~(\ref{eq:delta2}) in contrast to
Eq.~(\ref{eq:delta}), makes a great difference when it comes to finding the optimal state for state estimation. 

Notice that because of the constraint (\ref{eq:constraint3}), the sum of enumerators in Eq.~(\ref{eq:delta2}) is fixed. For a moment, let us 
consider only the situation when all terms in Eq.~(\ref{eq:delta2}) are positive. The fact that the denominator
increases with $j$, together with the fixed sum of enumerators makes it not desirable to have contribution from terms with large $j$. On 
the other hand, for high values of ${r}$ it impossible fulfill the constraint (\ref{eq:constraint3}) without such terms. As a result, 
intermediate $j$, which are as low as possible, yet high enough to fulfill (\ref{eq:constraint3}), will prove to be optimal.

Observe, that it is not desirable to have ${r}_j < 1$, at least for $j>s$. 
If for any $j^\prime>s$ we have ${r}_{j^\prime}<1$, then it is better to take ${r}_{j^\prime}=1$, decrease $p_{j^\prime}$, and
increase $p_{j^{\prime\prime}}$ for some $j^{\prime\prime}<j^\prime$. This will increase Eq.~(\ref{eq:delta2}) and at the same time
allows to keep the constraint (\ref{eq:constraint3}) satisfied.

Let us proof the following lemma, still assuming that terms in Eq.~(\ref{eq:delta2}) are positive:

\emph{Lemma.} If for a given ${r}$ the optimal fidelity correspond to $p_{j^\prime}=1$ for certain $j^\prime > s$, 
then after decreasing ${r}$ by a small amount $\epsilon$ ($0 \leq \epsilon \leq 2/N$), the optimal fidelity is achieved by taking the parameters $p_j$ of the form:
$p_{j^\prime}= 1 - \epsilon N/2$, $p_{j^\prime-1}=\epsilon N /2$, and the remaining $p_j$ equal zero.

\emph{Proof.}  
From Eq.~(\ref{eq:constraint1}) we have $j^\prime=N {r}/2$.
First we allow only one additional $p_{j^{\prime\prime}}$ ($j^{\prime\prime} \neq j^\prime$) to have a nonzero value.
 After decreasing ${r}$ by $\epsilon$, constraints are the following: $p_{j^\prime} + p_{j^{\prime\prime}}=1$, 
$p_{j^\prime} j^\prime + p_{j^{\prime\prime}} j^{\prime\prime} = N ({r} - \epsilon)/2$ and $j^{\prime\prime}<j^{\prime}$. The $\Delta_{\textrm{unknown}}$ now reads:
\begin{equation}
  \label{eq:25}
  \Delta_{\textrm{unknown}}=\frac{N \epsilon}{2(j^{\prime}-j^{\prime\prime})} \left(\frac{j^{\prime\prime}}{j^{\prime\prime}+1}- \frac{j^{\prime}}{j^{\prime}+1} \right)+\frac{j^{\prime}}{j^{\prime}+1}.
\end{equation}
The above formula attains the highest value for the highest possible $j^{\prime\prime}$ i.e. $j^{\prime\prime}=j^{\prime}-1$. 
Allowing only one additional $p_j^{\prime\prime}$ we find that the optimal case corresponds to $j^{\prime\prime}=j^{\prime}-1$. Thanks to the simple (linear in $p_j$) form of 
Eqs.~(\ref{eq:delta2},\ref{eq:constraint3}), fidelity for any other combination of $p_j$, with more than 
two nonzero terms, can be written as a convex combination of fidelities corresponding to case with only two non zero $p_j$.
Consequently, the optimal case is that with only two nonzero terms i.e. $p_{j^{\prime}}$, $p_{j^{\prime}-1}$. Using the constraint (\ref{eq:constraint3}) we arrive at:
$p_{j^{\prime}}= 1 - \epsilon N/2$, $p_{j^{\prime}-1}=\epsilon N/2$. The discussion is only valid for $\epsilon \leq 2/N$. For $\epsilon>2/N$, $p_{j^\prime}$ according to the derived formulas would be negative. $\blacksquare$

If ${r}=1$, there is no freedom for $p_j$ (\ref{eq:constraint3}), and 
we have to choose $p_{\mbox{\tiny{$N/2$}}}=1$. This choice corresponds to the product of pure states $\tilde{\rho}=|\psi\rangle\langle \psi|^{\otimes N}$. 
When we lower ${r}$ by a small amount ${r}=1- \epsilon$ ($0 <\epsilon \leq 2/N$), then using the lemma, the optimal state is
given by: $p_{\mbox{\tiny{$N/2$}}-1}= N(1-{r})/2$, $p_{\mbox{\tiny{$N/2$}}}=1-N(1-{r})/2$. For ${r}=1-2/N$ we obtain the situation in which only $p_{\mbox{\tiny{$N/2$}}-1}=1$ is nonzero, and we can again use the lemma.
We can carry on this process and find optimal states for smaller $r$ till we reach a state where $p_{s}=1$. 

It remains to discuss, whether negative terms in Eqs.~(\ref{eq:delta2},\ref{eq:constraint3}) can increase the fidelity. At first sight, it seems that they cannot.
Taking negative term, means that in order to fulfill (\ref{eq:constraint3}), we have to use positive terms with higher $j$, 
which will effectively make $\Delta$ smaller. 
This reasoning is not valid, however, for highly mixed local state (very low ${r}$). 

Consider first the case when $N$ is odd. If ${r}=1/N$, then according to the above discussion, the optimal state has $p_{1/2}=1$. For ${r}<1/N$ we cannot 
use the above scheme, since $j=1/2$ is the smallest $j$. One of the options is to shorten ${r}_{1/2}$ in order to fulfill (\ref{eq:constraint3}), but a better
strategy is to use a negative term with $j=N/2$, and keep ${r}_{1/2}=1$. Because of $j+1$ in the denominator of Eq.~(\ref{eq:delta2}), we gain more by keeping ${r}_{1/2}=1$, 
than we lose due to negative contribution from $j=N/2$ term. This also explains why we use $j=N/2$ and not a term with some other $j$ -- by doing so we subtract
as little as possible, as the denominator of $j=N/2$ term is the largest. 

For $N$ even, consider ${r}=2/N$, for which the optimal state corresponds to $p_{1}=1$. Lowering ${r}$, we can keep the constraint (\ref{eq:constraint3}) by
using $j=0$ term. Although this is the optimal strategy according to the lemma, it is no longer optimal if we allow negative terms. 
This is because, the term with $j=0$ is the only one which does not contribute to the fidelity at all. 
It is more advantageous to keep $p_{1}$ as large as possible, 
and fulfill the constraint (\ref{eq:constraint3}) by adding negative contribution from $j=N/2$, then we gain more thanks to a larger $p_{1}$ term than
we lose from subtracting $j=N/2$ term (again thanks to the denominator $j+1$).

Summarizing this lengthy reasoning, the optimal $N$--qubit state for state 
estimation with fixed local Bloch vector length ${r}$, when structure of correlations is not known, 
has all ${r}_j=1$ and coefficients $p_j$ read:
\begin{itemize}
\item{For $r \geq S/N$.

$
\begin{array}{l}
\left\{\begin{array}{lcl}
p_j&=&\dfrac{{r} N}{2}+1-j \\
 p_{j-1}&=&j-\dfrac{{r} N}{2}
\end{array}\right.
 \textrm{ when } j-1\leq  \dfrac{{r} N}{2} \leq j 
\end{array}
$
}
\item{For ${r} < S/N$.

$
\begin{array}{l}
\left\{\begin{array}{lcl}
p_{S/2}&=&\dfrac{({r}+1)N}{N+S} \\

 p_{N/2}&=&\dfrac{S-{r}N}{N+S} \quad \textrm{(antiparallel)} 
\end{array}\right.
\end{array}
$
 }
\end{itemize}
where $S=2$ for $N$ even, and $S=1$ for $N$ odd. 

With the above expressions one can easily calculate the optimal state and using Eq.~(\ref{eq:delta2}) the corresponding 
optimal fidelity of estimation of the optimal state. In the asymptotic limit $N \to \infty$, 
up to the leading order in $1/N$ the fidelity reads: $F=1-1/({r}N)$, and depends explicitely on ${r}$ in constrast to the case
when correlation structure is known.

\begin{figure}[t]
\begin{center}
\includegraphics[width=0.5\textwidth]{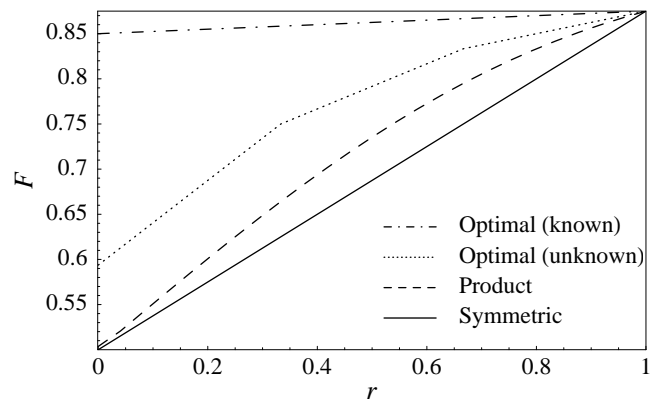}
\caption{The figure presents the comparison of the optimal fidelity of state estimation on four different $6$-qubit states:
the state supported on the symmetric subspace (solid curve), the product state $\rho^{\otimes N}$ (dashed curve), 
the state which we have found out is optimal from the point of view of state estimation when the structure of correlations is not known (dotted curve)
and the optimal state when the structure of correlations is known (dot-dash curve)
The goal of estimation is to estimate the direction of the Bloch vector of single particle reduced density matrices. The value of fidelity is plotted against 
the length of the Bloch vector ${r}$. For fixed ${r}$ the state supported on the symmetric subspace (the kind of state that comes out from the optimal cloning machine)
yields the lowest value of fidelity. }
 \label{fig:fid}
\end{center}
\end{figure}

In Fig.~\ref{fig:fid} we present the plots for the optimal estimation fidelity on the state of six qubits in four cases; when $\tilde{\rho}$ is:
$\tilde{\rho}_\textrm{sym}$ (supported on the symmetric subspace), 
 $\tilde{\rho}_\textrm{prod}$ (product state), $\tilde{\rho}_\textrm{unknown}$ (the optimal state for estimation when the structure of correlations is not known), 
$\tilde{\rho}_\textrm{known}$ (the optimal state for estimation when the structure of correlations is known), 
The length of the Bloch vector of the reduced single particle density matrix ${r}$ is depicted on the $x$ axis.
The edgy shape of the dotted line is due to  the fact, that in the case of unknown structure of correlations,
different pairs of $p_j$ are nonzero for different values of $r$.

The $\tilde{\rho}_\textrm{sym}$ state appears to be the worst among these four states, 
from the point of view of state estimation. Actually, it is the worst state in general 
for estimation, when the structure of correlation is known. This 
can be seen from the Eq.~(\ref{eq:fidknown}).
Only $p_{\mbox{\tiny{$N/2$}}}=1$ is non zero in this case, so only 
the term with the biggest denominator contribute. This makes the fidelity the smallest possible among all state with the same ${r}$.
This in particular proves the inequality $F_\textrm{prod} \geq F_\textrm{sym}$ mentioned in Sec.~\ref{optimal:estimation}. 

Notice also that $F_\textrm{known}$, and $F_\textrm{unknown}$ do not converge to $1/2$ as ${r} \to 0$, contrary to $F_\textrm{sym}$ and $F_\textrm{prod}$.
Even if local state are very mixed (arbitrary close to maximally mixed state), the estimation fidelity can be significantly higher than 
that $1/2$ (which corresponds to random guessing).

The state supported on the symmetric subspace $\tilde{\rho}_\textrm{sym}$, is interesting as this is the kind of state coming out from the optimal cloning machines. 
This is no coincidence that the state coming out from the optimal cloning machine is the worst from the point of view of state estimation. 
When we are doing $M \to N$ cloning, we can produce at the output only such a state of $N$ copies that can be estimated with
fidelity no higher than the optimal state estimation on $M$ pure state. Otherwise, we would violate state estimation bounds.
 In cloning we insist on the highest possible fidelity of clones. We attempt to obtain 
the highest possible value of ${r}$ (which determines local fidelity of clones). Doing so, and in order not to violate state estimation limits, 
we are forced to end up with the state which is the worst from the point of view of state estimation, among all the states with given ${r}$.

The problem we have considered here, carries some similarities to the problem of the optimal encoding and decoding of a spin direction 
\cite{gisin1999,bagan2001,bagan2001a,peres2000,chiribella2004}.
In the optimal encoding and decoding problem, one considers $N$ spins, and tries to find the optimal states of $N$ spins 
that would be optimal for the purpose of transmitting the information about a direction in space. The optimal state for this task
is in general ($N>2$ \cite{peres2000}) not a product states of $N$ spins, but a correlated state. This shows that the information can be better encoded in
correlated state than in a product one. Notice, however, that this result could not have been directly translated into our approach, as in our case we have 
strong restrictions on allowed states. We have considered permutationally invariant states, and the optimal state (class of states) for state estimation
that we have found, are states optimized under the constraint of fixed single particle density matrices.

We should also stress, that we have considered the problem of the estimation of the direction of the Bloch vector of the single-particle density 
and not the single-particle density matrix itself. In other words, we have not attempted to estimate the length of the Bloch vector, but 
rather assumed it is known. Hence, there remain an open problem of the optimal estimation on correlated copies of the single-particle density matrix,
when the length of the Bloch vector is unknown. This approach would be much more demanding, and in particular 
it is not clear what kind of distribution of correlated density matrices one should consider.
We leave this problem open, yet we conjecture that also in this approach one should arrive at the conclusion that the optimal state for state estimation
would be a correlated rather than a product state.
  
\section{Summary}
In this paper we have analyzed the problem of state estimation done on correlated copies in order to determine the direction of the Bloch vector of the
single-particle density matrix. In the case of qubits
we have found the optimal fidelity of estimation on $N$-qubit states which are permutationally invariant. The state which is the optimal from the point of
view of state estimation was derived. The optimal state is in general not a product state $\rho^{\otimes N}$, but has some correlations between qubits. 
As a result we have found out that the correlated state can be a better source of information about the single particle reduced density matrices than a product state.
In some cases, though, large correlations prevent us from extracting information about single particle reduced density matrices, as it is the case with clones coming out from 
the optimal cloning machine. One is not entitled, however, to claim in general that correlation always worsen our ability to extract information about 
single particle density matrix, as the optimal state for state estimation is in fact correlated.

\begin{acknowledgments}
I would like to thank Marek Ku\'s, for fruitful discussions and encouragement for this work. 
I am also grateful to Emil Bagan, John Calsamiglia and Ramon Mu\~noz-Tapia, 
for time and effort they spent on critical reading of the paper, spotting and helping to correct errors.
This work was supported
by the EC grant QUPRODIS, contract No IST-2001-38877 and the Polish Ministry of
Scientific Research and Information Technology under the (solicited) grant No
PBZ-Min-008/P03/03.
\end{acknowledgments}

\end{document}